\documentclass[referee]{raa}           
\usepackage{graphicx,times}
\usepackage{natbib}
\usepackage{amssymb,amsmath}
\bibpunct{(}{)}{;}{a}{}{,}

\begin{document}

   \title{Investigation of two coronal mass ejections from circular ribbon source region: Origin, Sun-Earth propagation and Geo-effectiveness
}

 \volnopage{ {\bf 20XX} Vol.\ {\bf X} No. {\bf XX}, 000--000}
   \setcounter{page}{1}

   \author{Syed Ibrahim\inst{1}, Wahab Uddin\inst{1}, Bhuwan Joshi\inst{2}, Ramesh Chandra\inst{3}, Arun Kumar Awasthi\inst{4}
      }


   \institute{ Aryabhatta Research Institute of Observational Sciences, Nainital 263002, India; {\it syed@aries.res.in}\\
        \and
             Udaipur Solar Observatory, Physical Research Laboratory, Udaipur 313004, India\\
	\and
Department of Physics, DSB Campus, Kumaun University, Nainital 263002, India\\
\and 
CAS Key Laboratory of Geospace Environment, Department of Geophysics and Planetary Sciences, University of Science and Technology of China, Hefei 230026, China\\
\vs \no
   {\small Received 20XX Month Day; accepted 20XX Month Day}
}

\abstract{In this article, we compare the properties of two coronal mass ejections (CMEs) that show similar source region characteristics but different evolutionary behavior in the later phases. We discuss the two events  in terms of their near-Sun characteristics, interplanetary evolution, and geo-effectiveness. We carefully analyzed the initiation and propagation parameters of these events to establish the precise CME-ICME connection and their near-Earth consequences. The First event was associated with poor geo-magentic storm disturbance index (Dst $\approx$-20 nT) while the second event is associated with intense geomagnetic storm of DST $\approx$-119 nT. The configuration of the sunspots in the active regions and their evolution are observed by Helioseismic and Magnetic Imager (HMI). For source region imaging, we rely on data obtained from Atmospheric Imaging Assembly (AIA) on board Solar Dynamics Observatory (SDO) and H$\alpha$ filtergrams from Solar Tower Telescope at Aryabhatta Research Institute of Observational Sciences (ARIES). For both the CMEs, flux rope eruptions from the source region triggered flares of similar intensities ($\approx$M1). At the solar source region of the eruptions, we observed circular ribbon flare (CRF) for both the cases, suggesting fan-spine magnetic configuration in the active region corona. The multi-channel SDO observations confirm that the eruptive flares and subsequent CMEs were intimately related to the filament eruption. Within the field of views of Large Angle Spectrometric Coronograph (LASCO) field of view (FOV) the two CMEs propagated with linear speeds of 671 and 631 km s$^{-1}$, respectively. These CMEs were tracked up to the Earth by Solar Terrestrial and Relational Observatory (STEREO) instruments. We find that the source region evolution of CMEs, guided by the large-scale coronal magnetic field configuration, along with near-Sun propagation characteristics, such as, CME-CME interactions, played important roles in deciding the evolution of CMEs in the interplanetary medium and subsequently their geo-effectiveness.
\keywords{Coronal Mass Ejections, Flare, Interplanetary Coronal Mass Ejections
}
}

   \authorrunning{Syed Ibrahim et al. }            
   \titlerunning{ Investigation of two coronal mass ejections from circular ribbon source region}  
   \maketitle

%
\section{Introduction}           
\label{sect:intro}

Flares and coronal mass ejections (CMEs) are the largest eruptive phenomena in our solar system. Flares are characterized by a sudden catastrophic release of energy in the solar atmosphere. Huge amount of energy (in excess of $10^{32}$ erg) is released within a few of minutes \citep[see review by][]{Benz2017}. The standard flare model provides a basic framework to understand the relationship between solar flares and filament eruptions \citep[see e.g.,][]{Car1964, Stur1966, Hira1974, Kopp1976}. 

In the standard flare model, the expansion of a filament from the core of the AR forms an integral part of the triggering process of an eruptive flare \citep{Shibata1999, Joshi2012}. Filaments are the large magnetic structures which are capable to store magnetic energy to drive the eruptions \citep[see e.g., ][]{Gibson2006, Krall2007, Wahab2012, Mitra2018}. The overlying AR magnetic loops are stretched by the filament eruption and the magnetic reconnection sets in underneath the erupting filament. The accelerated particles precipitate along the newly formed post-reconnected field lines and subsequently produce the bright flare ribbons in chromosphere \citep[see e.g., ][]{Demo1996, Joshin2015, Joshi2017a}. Flare ribbons are observed in the various geometric shapes like parallel, J-shaped, X-shaped, circular and multi ribbons \citep[see e.g., ][]{Jan2016, Li2016, Masson2009, Reid2012, Kushwaha2014, Joshi2017b, Zhong2019, Tor2019, Pooja2020, Liu2020}. Chromospheric flare ribbons are generally situated at locations intersected by separatrices dividing domains of distinct connectivity or quasi-separatrix layers possessing strong connectivity gradients \citep{Hudson2011}. Magnetic field lines associated with a 3D coronal null point usually display a fan–spine configuration, where the dome-shaped fan portrays the closed separatrix surface and the inner and outer spine field lines in different connectivity domains pass through the null point \citep{Demo1994, Demo1997}. The footpoint of the inner spine has a magnetic polarity opposite to those of the fan, which forms a circular polarity inversion line \citep{Lau1990, Aulanier2005}. Magnetic reconnection can be induced in such single null points, as the fan/spines deviate from the null when subject to shearing or rotational perturbations \citep{Pontin2007, Pontin2007a, Qiu2007}. It is then expected that as a result of null-point reconnection, flare emissions at the footpoints of the fan field lines would constitute a closed circular flare ribbon, and that the spine-related flare footpoint would be a compact source.

The eruptive filaments, if leaves the corona successfully against the constraining force of the overlying magnetic fields and gravity of the Sun, forms the integral part of a CME and its inner most part is identified as core of the CME \citep{Suraj2020, Prabir2020, Gopal2020}. CMEs erupt from the lower solar atmosphere and propagate into the interplanetary space through solar corona with various velocity range from hundreds of km $s^{-1}$ to 3000 km $s^{-1}$ \citep{Gopal2000,Schmieder2015, Syed2019, Gopal2020}. In the near-Sun region, it is not completely understood how the magnetic reconnection and geometry/topology of overlying magnetic fields contribute in the kinematics of the CMEs \citep[see e.g.,][]{Wahab2012,Joshi2013,Kushwaha2015,Joshi2016, Chandra2017,Mitra2019,Anitha2020}.

The kinematic evolution of CMEs continuously change from near-Sun region to near Earth interplanetary medium. Geomagnetic storms are associated with CMEs and their interplanetary counterparts \citep{Zurbuchen2006,Zhao2014,Kilpua2017,Alex2018,Bravo2019,Hema2021}. To understand the complex process associated with CME origin and its evolution in the interplanetary medium, it is necessary to connect the near-Sun observations, interplanetary radio emissions, and in-situ measurements at 1 AU \citep[see e.g.,][]{Joshi2018}. In general, ICMEs acceleration and deceleration depends on their speed relative to the solar wind speed \citep{Manoharan2010, Shanmug2014, Shanmug2015, Joshi2018}. Slow CMEs are accelerated by the solar wind while the fast CMEs are decelerated by the solar wind. Therefore, the transit time of CMEs depends on the state of ambient solar wind condition as well as the CME-CME/CME-solar wind interactions \citep[see, e.g.,][]{Manoharan2004, Manoharan2006, Vrsnak2007, Gopal2015, Syed2015,Sudar2016, Syed2017}. Understanding the CME kinematics and its propagation are very important in the concept of interplanetary space weather. Notably, front-sided Earth directed high speed halo CMEs are potential candidates for major geomagnetic storms. This paper is an attempt to explore the CME initiation and propagation characteristics for a special category of flaring event, called circular ribbon eruptive flares, by utilizing the multi-wavelength and multi-instrumental measurements.

In this study, we discuss two Sun-Earth connecting geo-effective CMEs that erupted on 2013 May 02 and 2014 February 16. The events were originated from the AR NOAA 11731 and 11977 respectively. The source region characteristic of the two events are identical but the subsequent CME evolution shows differences. Hence, it is worth to compare the two events in terms of their near-Sun properties, interplanetary evolution, and near-Earth consequences. A very interesting aspect of the source region of these CMEs was their association with impulsive solar eruptive circular ribbon flares of similar intensity class ($\approx$M1). Notably, circular ribbon flares are generated in active regions that exhibit a particular magnetic topology, known as the fan-spine configuration, which is different from the classical magnetic field configuration of the flaring region where the 'standard' eruptive flares originate \citep[see e.g.,][]{Vrsnak2003, Long2005}. The investigations concerning the CME origin and the Sun-Earth connection associated with eruptive circular ribbon flares are still very limited; the present study makes inroads toward such objectives. We used multi-wavelength and multi-point observations to connect the CME evolution in the near-Sun region to the near Earth space. These solar eruptions lead geomagnetic storm with Dst values from moderate (-20 nT) to intense (-119 nT). In Section~\ref{sec:data}, we provide details about the various observational resources and data analysis. The characteristics of the flares, CMEs, ICMEs and their inter connection are given in Section~\ref{sec:analysis_results}. The summary of results and conclusions are given in the final Section~\ref{sec:conclusions}.


\section{Data and Selection Criterion}
\label{sec:data}
In this paper, we used following data sources to analyze the eruptive flares and CMEs/ICMEs in the Sun-Earth connecting space:

To understand the source region of CMES, data are obtained from Atmospheric Imaging Assembly \citep[AIA;][]{Lemen2012} and Helioseismic Magnetic Imager \citep[HMI;][]{Schou2012} on board the Solar Dynamics Observatory \citep[SDO;][]{Pesnell2012}. We have analyzed EUV images of the Sun taken in 94~\AA, 171~\AA, 193~\AA~and 304~\AA~filter of AIA which provide useful information about the magnetic loops connectivities, chromospheric features and flare related high-temperature structures. The white light and magnetogram images from HMI provide the information about photospheric sunspot evolution and magnetic field of AR.

We used 15 cm f/15 solar tower telescope H$\alpha$ observations on 2013 May 02 from Aryabhatta Research Institute of Observational Sciences (ARIES), Nainital, India. The image size was enlarged by a factor of two using a Barlow lens. The images were recorded by a 16 bit 1K x 1K pixel CCD camera system having a pixel size of 13  $\mu^{2}$. The cadence for the images is 2 - 5 seconds. 

Ground based radio spectrograph is used to study the radio wave mechanism of plasma in near-Sun. For this purpose, data are obtained from extended Compact Astronomical Low Cost Frequency Instrument for Spectroscopy and Transportable Observatory (e-CALLISTO; \cite{Benz2005}). Flare details in soft X-ray (SXR) channels are obtained from Geostationary Operational Environmental Satellite (GOES) observations.

To study the CME dynamics in the near-Sun region, we have used white light images from the Large Angle Spectrometric Coronagraph \citep[LASCO;][]{Brueckner1995} on board of Solar and Heliospheric Observatory (SOHO). For understanding the CME propagation in the interplanetary space, we used Solar Terrestrial and Relational Observatory (STEREO) observations. For the ICME tracking, we used four instruments:  inner coronagraph (COR1), outer coronagraph (COR2) and two heliospheric imagers (HI 1\& 2). Using these instruments one can track the CMEs/ICMEs transit from near Sun to beyond the Earth \citep{Kaiser2008}.

We have analyzed in-situ interplanetary solar wind plasma and magnetic field parameters associated with the CMEs using multi-instrument data which are collectively available at the Coordinated Data Analysis Web (CDAWeb\footnote{https://cdaweb.sci.gsfc.nasa.gov/index.html/}).

\section{Analysis and Results}
\label{sec:analysis_results}

\subsection{Flaring Active Regions NOAA 11731, 11977 and Source of Circular Ribbon Flares}
\begin{figure*}
\centering
\includegraphics[width=1.0\textwidth]{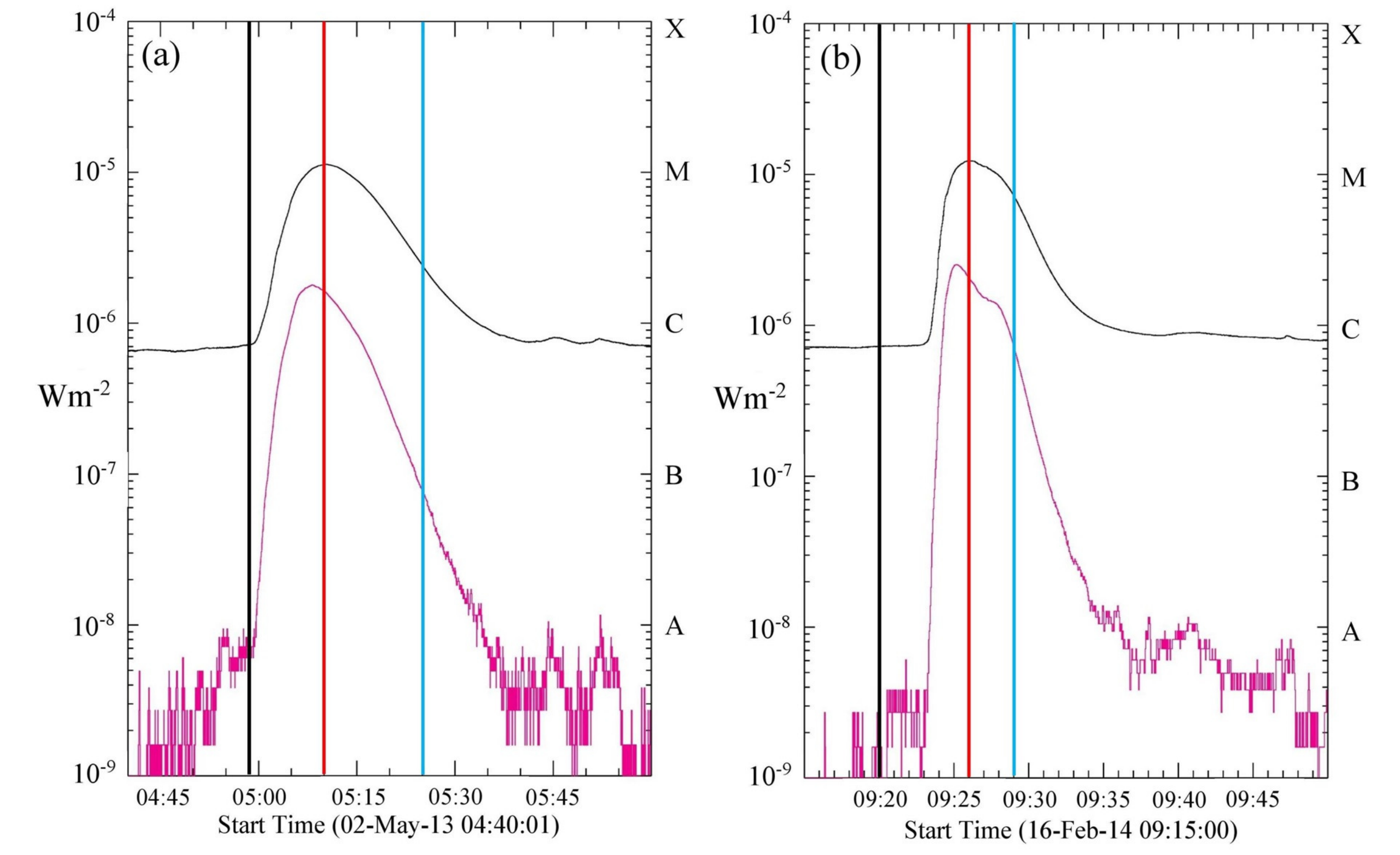}
\caption{GOES light curves showing the M1.1 class flare evolution observed on (a) 2013 May 02 04:45 - 05:45 UT and (b) 2014 February 16 09:18 - 09:36 UT. Two different wavelengths of 1-8 and 0.5-4~{\AA} correspond to disk-integrated X-ray emission in 1.5--12.5 keV and 3--25 keV energy range, respectively. Vertical solid lines of black, red and green colors indicate the flare start, peak and end times respectively.}
\label{GOES_CRF2}
\end{figure*}

In Figure~\ref{GOES_CRF2}, we show the plots for the evolution of two SXR flares of class M1.1 observed by GOES between 04:58--05:10 UT on 2013 May 02 and 09:20--09:29 UT on 2014 February 16. In these figures, we can see the SXR flux variation in the two different channels (0.5-4~\AA~and 1-8~\AA~respectively). The first flaring event, AR NOAA 11731 occurred on 2013 April 24. Initially AR displayed a $\beta\gamma\delta$ type magnetic configuration, then four days later it turned into $\beta\gamma$ type configuration. Then again it attained $\beta\gamma\delta$ configuration on 2013 May 02. The M1.1 class eruptive flare originated on 2013 May 02 when the location of the AR was N10W25.

\begin{figure*}
\centering
\includegraphics[width=1.0\textwidth]{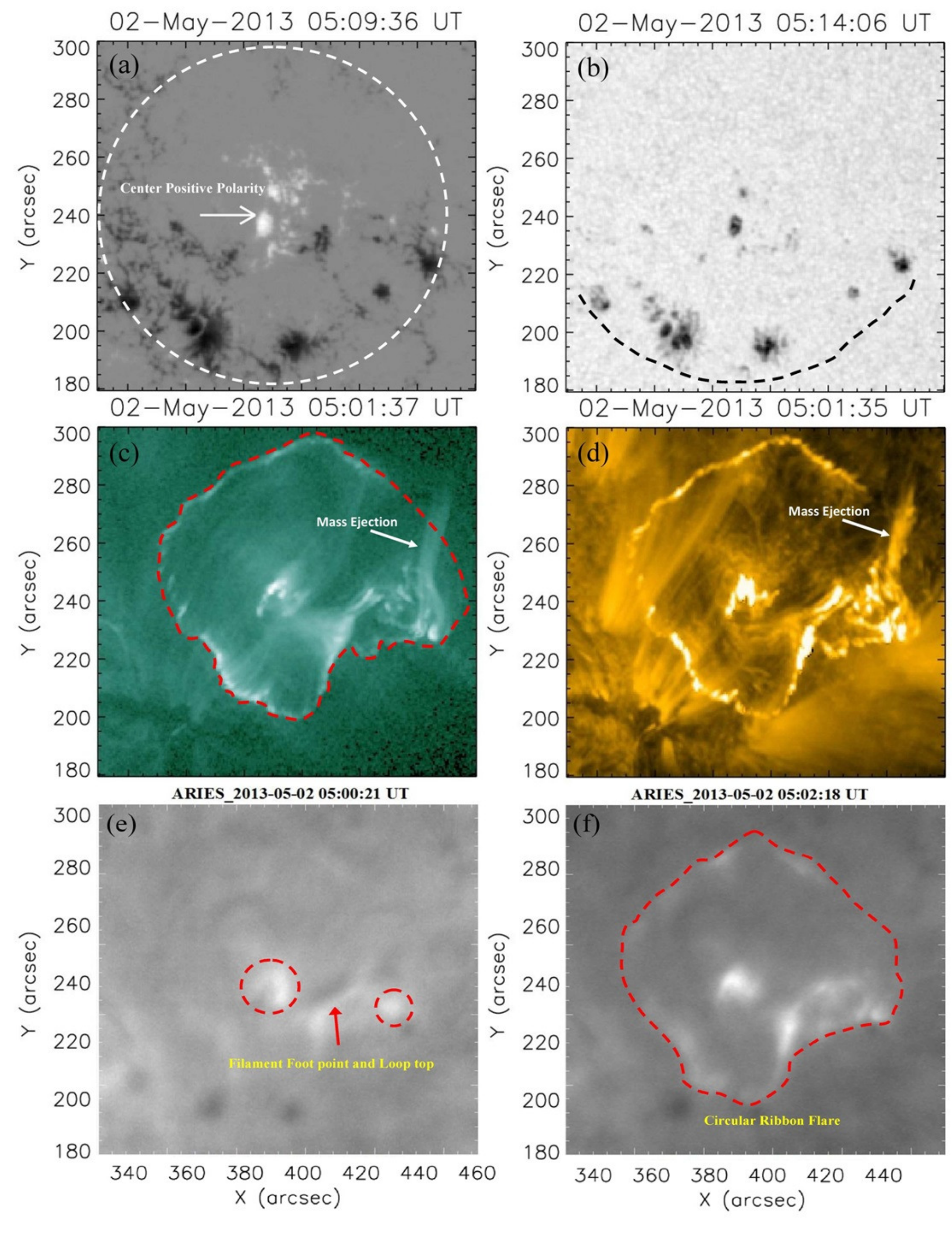}
\caption{(a) HMI  magnetogram image showing the central positive polarity surrounded by circular type of negative magnetic polarity region which is indicated by the dashed white circle. (b) HMI white light image showing the sunspot distribution in the AR where semi circular structure in the distribution of sunspots is marked by the dashed line. View of circular ribbon flare (c) in AIA~94~{\AA} and (d)~171~{\AA} observations. White arrow mark indicate the mass eruption from the AR. (e) ARIES solar tower telescope H$\alpha$ observation for the circular ribbon flare on 2013 May 02. Filament food points and loop top are indicated by the dashed red circles and arrow. (f) View of circular ribbon flare indicated by the dashed red line.
}
\label{M_andH_al_CRF2}
\end{figure*}

We present the magnetic configuration of AR NOAA 11731 in Figure~\ref{M_andH_al_CRF2}a. The white light sunspot image from HMI/SDO clearly indicates the leading part of the AR (see Figure ~\ref{M_andH_al_CRF2}b). The distribution of magnetic polarities in the leading part of AR forms approximately a semi circle which is indicated by the dashed line. There is a small filament that existed at the flaring location in the leading part of the active region which is surrounded by bright plages (see Figure ~\ref{M_andH_al_CRF2}e). To compare the photospheric layer with the upper layers, we have shown AIA~94~\AA~and~171~\AA~images of the AR in panels c and d. Comparisons of ~94~\AA~and~171~\AA~filtergram with HMI magnetogram images clearly show that the region of mass eruption and CRF structure extend over the west side of the AR. The filament is connecting the positive and negative polarity of the leading sunspot group. Also the ARIES H$\alpha$ images clearly show the structure of CRF. In Figure \ref{M_andH_al_CRF2}e and f, initial filament is indicated by the red arrow mark which was observed on 2013 May 02 arround 4:59 UT. Filament dark structure and bright foot points are noted by red arrow and dashed red circle respectively.

The second flaring region NOAA 11977 was visible on the solar disk on 2014 February 10. Initially AR was a simple $\alpha$ type magnetic configuration, then three days later it turned into $\beta$ type configuration. The AR size as well as magnetic complexity developed and changed to a $\beta\gamma$ configuration on 2014 February 14. The intense geo-effective eruptive flare event originated on 2014 February 16 at location S10W00.

\begin{figure*}
\centering
\includegraphics[width=1.0\textwidth]{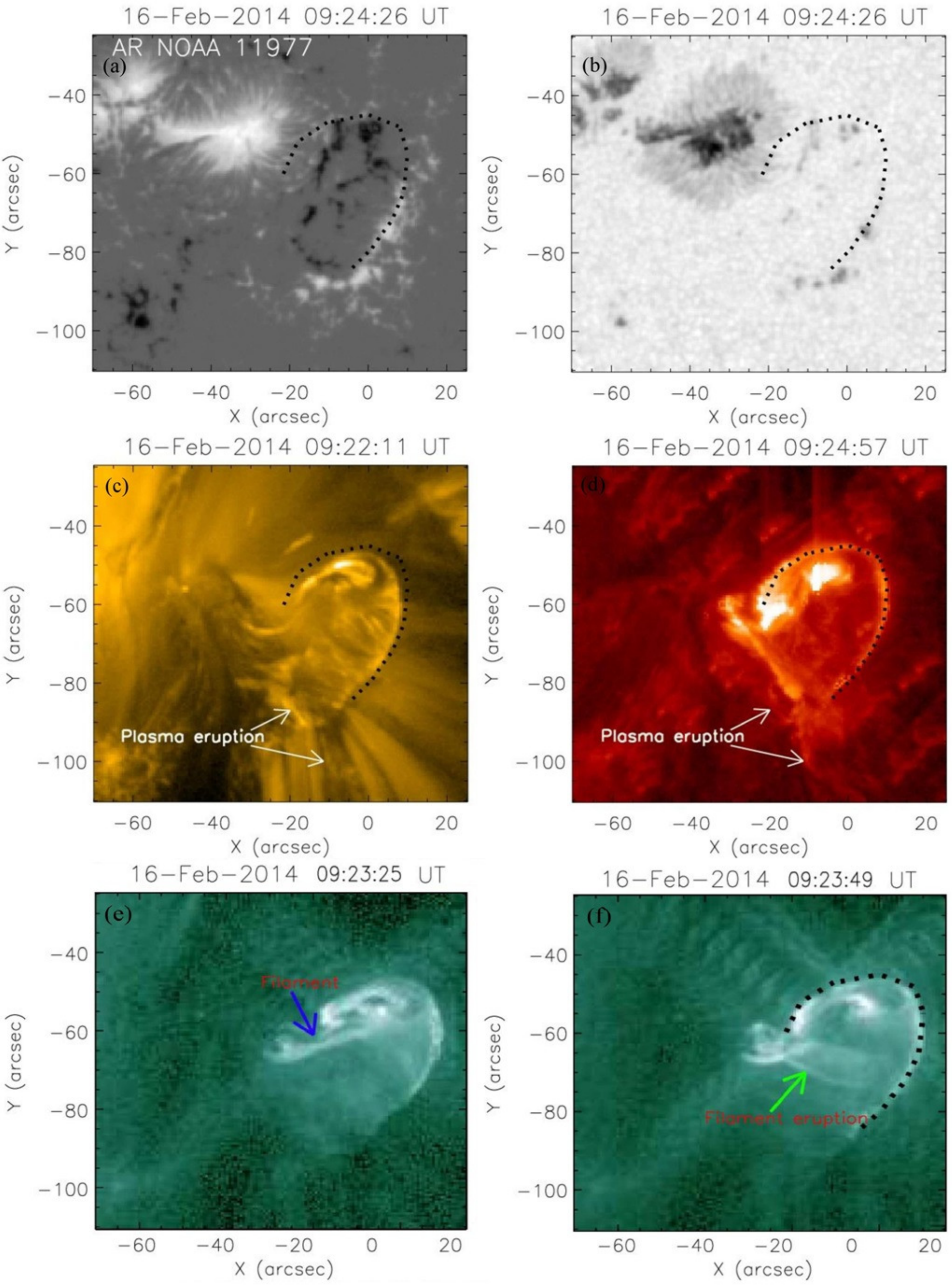}
\caption{Multi-wavelength view of AR NOAA 11977 during the flare peak time. (a) HMI  magnetogram and (b) white light images are covering the flare region. Circular ribbon flaring region is shown by the dotted curved lines. Circular ribbon flare is shown in AIA observations: (c) ~171~{\AA}, (d)~304~{\AA} channels. Filament and it erution signatures are very clear in the~94~{\AA} observations (see last two panels (e) and (f)). White arrow marks indicate the filament eruption/motion. The filament eruption is marked by the color arrow symbols in the last two panels.
}
\label{Multi_wave}
\end{figure*}

Figure~\ref{Multi_wave} shows the multi wavelength view of the AR NOAA 11977. The sunspot magnetogram and corresponding white light images (Figure \ref{Multi_wave}a and b) clearly indicate the leading and following parts of the active region. The leading sunspot group largely consists of some mixed polarity region at the edge of active region. A small filament erupted from the complex polarity region and caused the M1.1 impulsive flare and was associated with the moderated speed HALO CME. Finally, this CME produced a major geo-effective event. It reaches the Earth on 2014 February 19. 

During the flare evolution, clearly we can see that the twisted filament erupted from the complex polarity region and it produced circular ribbon flares (see Figure \ref{Multi_wave}c and d). Color arrow mark indicates the filament eruption from the merged polarity region (last two panels e and f). One leg of the filament is anchored on the strong sunspot magnetic field region and another leg anchored in the weak magnetic field region. The eruption was started from the weak field region. 

\subsection{Radio observations}

\begin{figure*}
\centering
\includegraphics[width=1.0\textwidth]{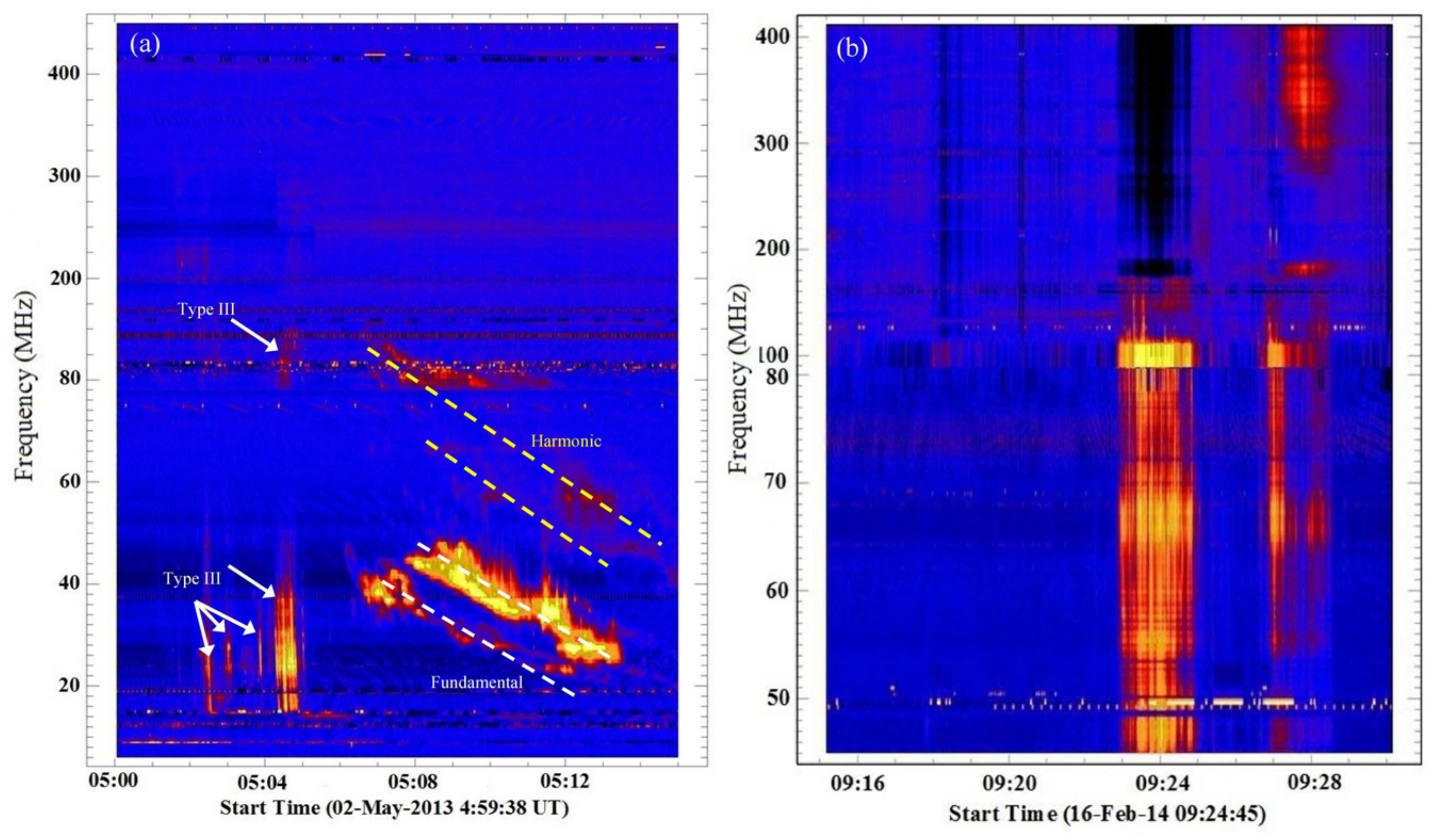}
\caption{(a) Dynamic radio spectrum showing the metric type III and type II corresponding to the flare of class M1.1 occurred on 2013 May 02. Upper and lower type II radio burst, fundamental and harmonic frequency band splitting are shown by the dashed white and yellow lines respectively. (b) Dynamic radio spectrum showing metric type III observation corresponding to the flare of class M1.1 occurred on 2014 February 16.} 
\label{CALLISTO_CRF2}
\end{figure*}

During these CRFs, e-CALLISTO spectrograph (Benz et al. 2005) observed emission at radio frequencies. The dynamic solar radio spectra are shown in Figures \ref{CALLISTO_CRF2} at a wide frequency range of 16 to 460 MHz. The dynamic spectrum of the first event displayed a type III burst  (Figure \ref{CALLISTO_CRF2}a). At the peak time of the flare (around 05:05 UT), we observed the type III burst in the frequency range of 150 MHz to 16 MHz. Type III burst implies the opening of the magnetic filed lines and subsequent ejection of relativistic electron beam \citep{Joshi2018}. We can see that a type II radio burst is followed by the type III burst. In Figure \ref{CALLISTO_CRF2}a, the type II fundamental and hormonic band splitting pattern is indicated by the dashed white and yellow lines respectively. From the metric type II radio observations, we estimate the shock formation height range for upper and lower frequency ranges (120-40 MHz and 46-25 MHz). Corresponding height ranges for the above frequencies are 1.18-1.62R$_{\odot}$ and 1.55-1.91R$_{\odot}$ (for upper and lower band) respectively. Also the corresponding shock speed is estimated to be 728 - 835 km $s^{-1}$ using Newkirk one fold density model \citep{Newkirk1961} which is well applicable for the lower heights of the solar atmosphere. The average shock speed (782 km $s^{-1}$) is roughly comparable to the CME speed (671 km $s^{-1}$) within the LASCO FOV. 

The dynamic radio spectrum of the second flare is given in Figure \ref{CALLISTO_CRF2}b. At the peak time of the flare around 09:24 UT, we observed the type III burst between the frequency range of 100 MHz to 45 MHz. There was no type II association, because the flare was associated with a moderate speed CME.

\subsection{CMEs characteristics}
\begin{figure*}
\centering
\includegraphics[width=0.75\textwidth]{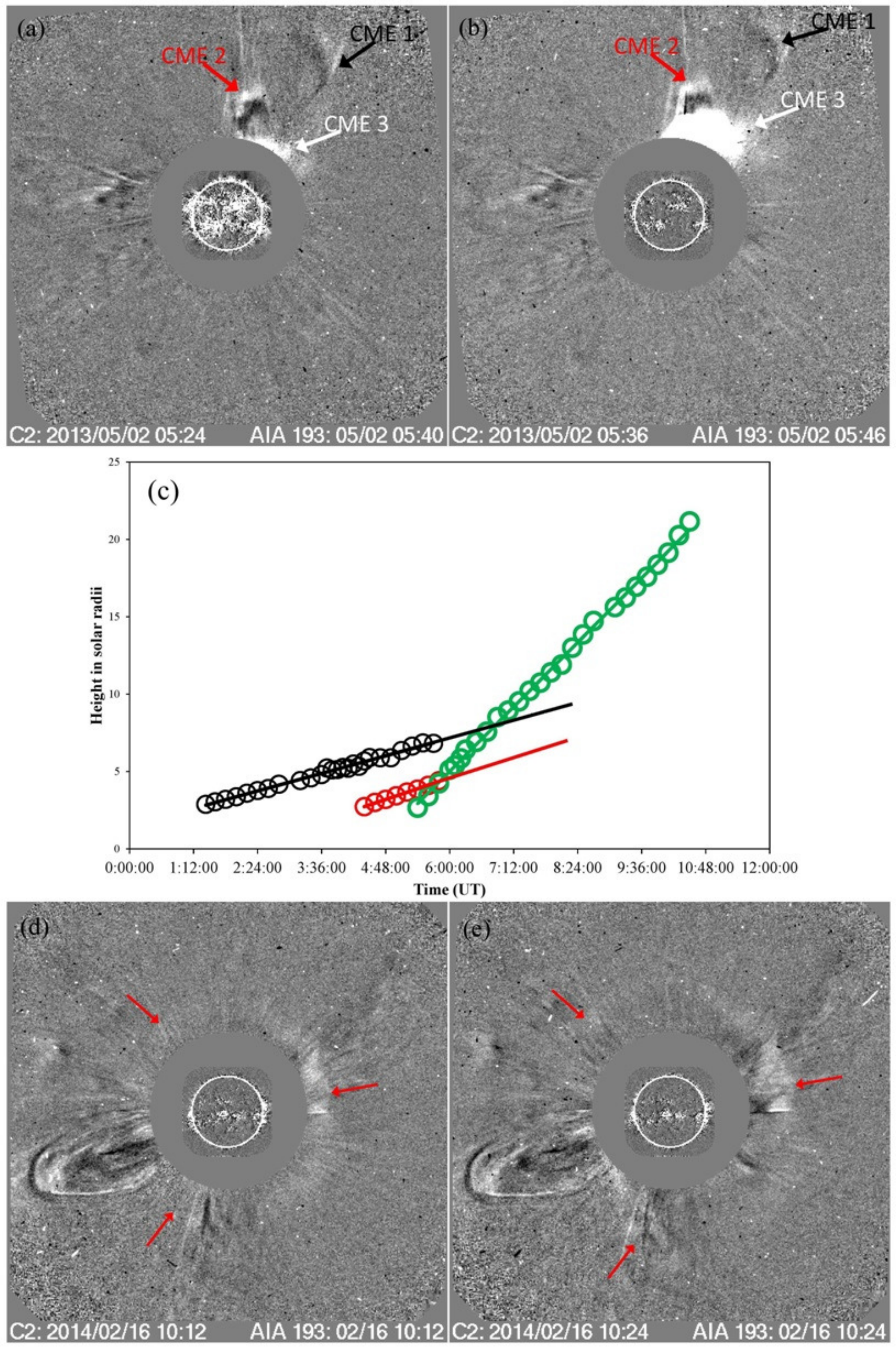}
\caption{Running difference images are derived from LASCO C2 observations (panels (a) and (b)) showing the propagation of the moderate speed CME originated from the disk center AR NOAA 11977 on 2013 May 02. The red and black arrows indicate the two pre-CMEs while the main CME is indicated by white arrow. (c) The three events interacted in the LASCO FOV. Similarly, last two panels are showing the propagation of the halo CME originated from the AR NOAA 11977 on 2014 February 16. The red color arrows indicate the halo CME propagation in the LASCO FOV.} 
\label{LASCO_CME_CRF2}
\end{figure*}

First flare is associated with a moderate speed CME, Figure \ref{LASCO_CME_CRF2}a and b shows that the white-light running difference images from LASCO C2 observations to show the CME observed on 2013 May 02. The CME first appeared at 2.66R$_{\odot}$ around 05:24 UT in the C2 FOV, and its linear speed was 671 km $s^{-1}$. Finally, the same CME tracked by the LASCO C3 coronagraph at 21.15R$_{\odot}$ around 10:30 UT. The main CME interacted with two pre-CMEs in the LASCO FOV. These two pre-CMEs were consecutively erupted around 01:25 UT and 04:24 UT (see Figure \ref{LASCO_CME_CRF2}c). These events were associated with lesser angular widths, also they propagated in the same direction of main CME which is erupted around 05:24 UT. After the eruption of main CME, we can see the clear interaction of these events. Central and mean position angle of these three CMEs are nearly same around 353 degree.

Second flare is also associated with moderate speed halo CME, Figure \ref{LASCO_CME_CRF2}d and e shows that the white-light images of LASCO C2 coronagraph observations on 16 February 2014. CME onset was observed at 2.55R$_{\odot}$ around 10:00 UT in the C2 FOV, and its linear speed in the LASCO field of view was 634 km $s^{-1}$. Finally, the same CME tracked by the LASCO C3 coronagraph at 6.77R$_{\odot}$ around 11:18 UT. This CME event caused an intense storm at the near-Earth environment. Immediately, after the main geo-effective CME, a strong bright CME was ejected from the eastern part of the solar limb, so that we can not track the main geo-effective CME in the LASCO FOV after the distance of 7R$_{\odot}$. In the second event, we do not find any evidence of the interaction between the pre-post CMEs. This geo-effective CME was observed by the STEREO coronagraph from Sun to Earth. 

Further, we calculated the CME 3D speed using STEREO and LASCO observations for both the events. From the Graduated Cylindrical Shell (GCS) Model, the obtained speed values for the first and second CMEs are 641 km $s^{-1}$ and 690 km $s^{-1}$, respectively \citep{Thernisien2006, Singh2018}.

\subsection{EUV waves and direction of CME propagation}

\begin{figure*}
\centering
\includegraphics[width=0.7\textwidth]{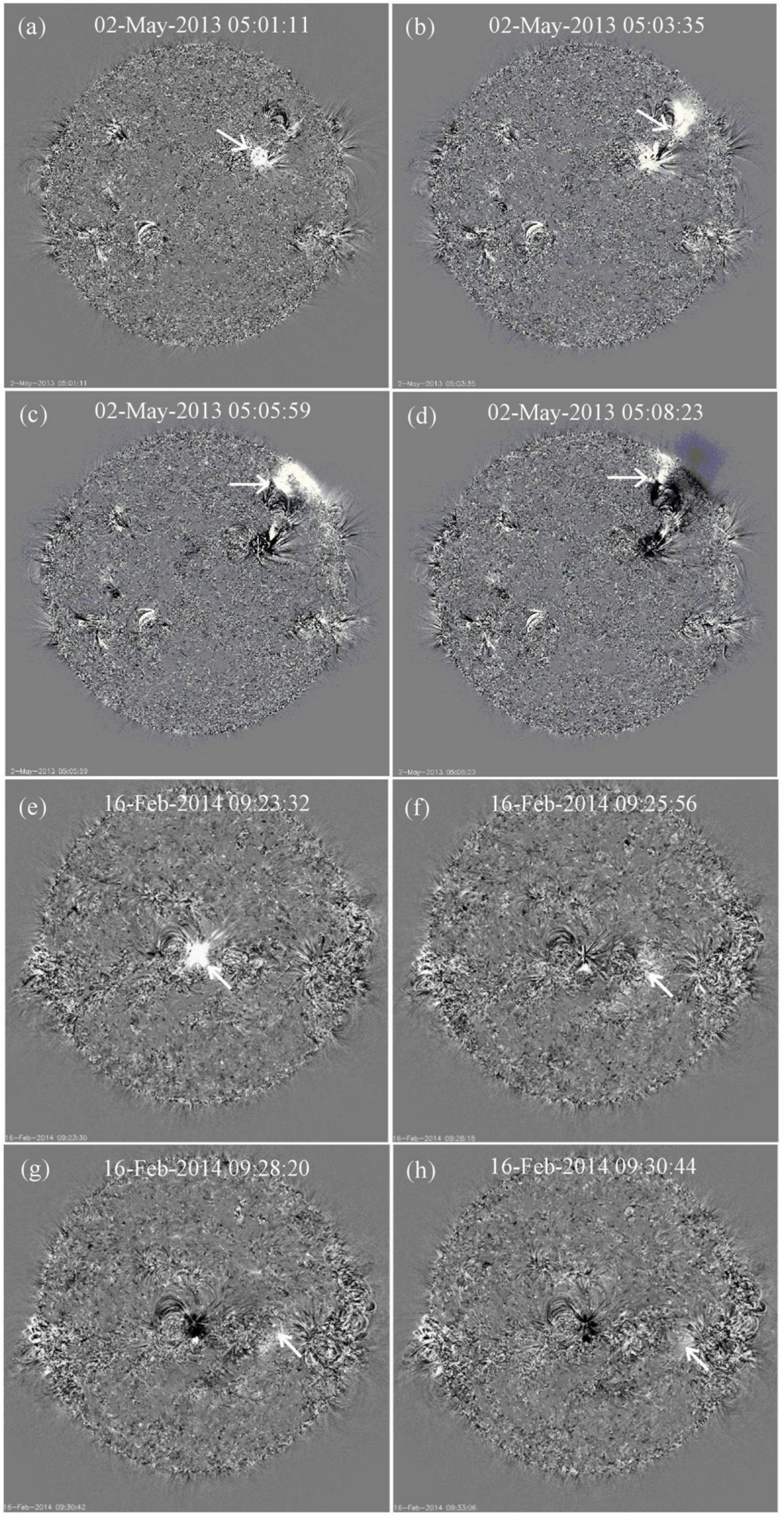}
\caption{The first CME event is associated with an EUV wave which is clearly seen in AIA difference images. Importantly, the EUV wave is deflected from the source region. The deflection of the EUV wave is indicated by the white arrows (panels a, b, c and d). There was no deflection in the case of EUV wave associated with the second event (see panels e, f, g and h). Above images are taken from the online available movies (www.lmsal.com/nitta).} 
\label{Deflection}
\end{figure*}

The geo-effectiveness of the CMEs depends upon its direction towards Earth. If the eruption is deflected away from the Earth direction, it can be less geo-effective. Therefore in this section, we are discussing about the direction of CME propagation. Both the eruption are associated with the EUV waves (Figure \ref{Deflection}). Since EUV waves are associated with solar eruptions, therefore the deflection in EUV wave provides the information about the deflection in CMEs. The deflection in EUV wave were observed in past \citep[see e.g.,][]{Pat2009, Filippov2010, Zuccarello2017, Ramesh2018}. The first wave is stronger and clearly associated with type II radio burst. However, we observed the deflection of this wave. For the second event, we do not find any deflection of the EUV wave and the CME comes directly towards the Earth. From the Figure \ref{Deflection}, we found that the first CME was deflected from the source region. As we can see, the eruption is not along the radial direction. Huge portion of the CME was moved to the north-east part of the Sun (see panels a, b, c and d). Finally, small part of the CME propagates in the interplanetary space. Mostly the low-latitude coronal holes are appearing frequently in the solar disk, so CME deflection by such coronal holes becomes very important \citep{Gopalswamy2009, Mohamed2012, Makela2013}. The deflections are thought to be caused by the magnetic pressure gradient between the eruption regions and the coronal holes \citep{Gui2001, Shen2011, Gui2011}. However, there are other processes that can significantly affect CMEs propagation, for example CME-CME interaction and CME deflection by large scale structures such as streamers \citep[see e.g.,][]{Temmer2012, Temmer2014, Gopalswamy2009, Wood2012, Kay2013, Panasenco2013, Gopalswamy2014}. The deflection angle of the part of the CME's non-radial propagation was much larger than we expect. In most of the reported cases, coronal longitudinal deflections are lesser than $\approx$20$^{\circ}$ \citep{Isavnin2014} while in this study, we found that the deflection angle of the CME is nearly $\approx$40$^{\circ}$.

Notably, there was no deflection in the case of the second CME, so that most of the CME portion reached the Earth (see Figure \ref{Deflection}e, f, g and h). For the first event, the deflection of CME within the source region is the primary cause for its lower geo-effectiveness (-20 nT). On the other hand, for the second event, bulk of the CME structure from the source region constituted the corresponding ICMEs and, hence, the event produced strong geo-effectiveness of -119 nT.

\subsection{ICMEs characteristics}

\begin{figure*}
\centering
\includegraphics[width=0.75\textwidth]{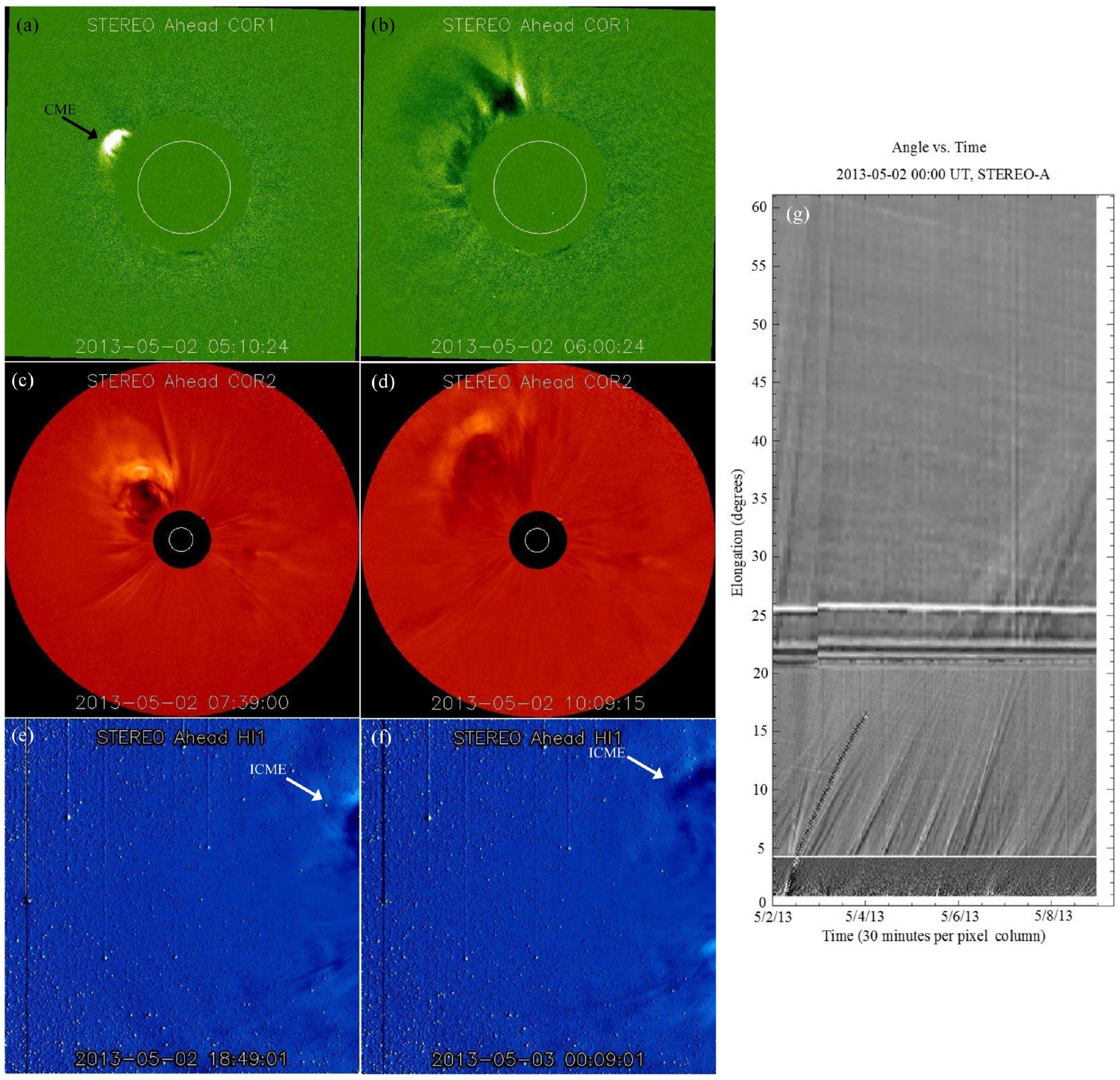}
\caption{CME/ICME observation in the STEREO FOV (for the corresponding CME see Figure \ref{LASCO_CME_CRF2}a and b). The evolution of the CME in (panels (a) and (b)) COR1 FOV, and the same CME in (panels (c), (d), (e) and (f)) COR2 and HI1  FOV. The propagation of the CME/ICME is clearly visible in the images. Portion of the CME is indicated by the arrows. (g) Time-elongation map (J-map) constructed using the STEREO/SECCHI spacecraft observations during the interval of 05 to 08 May 2013.} 
\label{stereo_CME_CRF2}
\end{figure*}

Direct evidence of CMEs propagation and interaction in the interplanetary medium is gathered from STEREO measurements using images from the coronagarph observations  \citep{Lugaz2012}. The evolution of the first CME with COR2 images and HI running-difference images are shown in Figure \ref{stereo_CME_CRF2}. Using these snapshot images, we clearly see that the CME propagation in the inner and outer heliosphere. CME initial positions are indicated by the black arrow mark in STA COR1 FOV (see Figure \ref{stereo_CME_CRF2}a and b). This particular CME observed in the STA around 10:24 UT. Further, interplanetary CME progation shown in Figure \ref{stereo_CME_CRF2}c, d, e and f by white arrow mark.

\begin{figure*}
  \centering
  \includegraphics[width=0.75\textwidth]{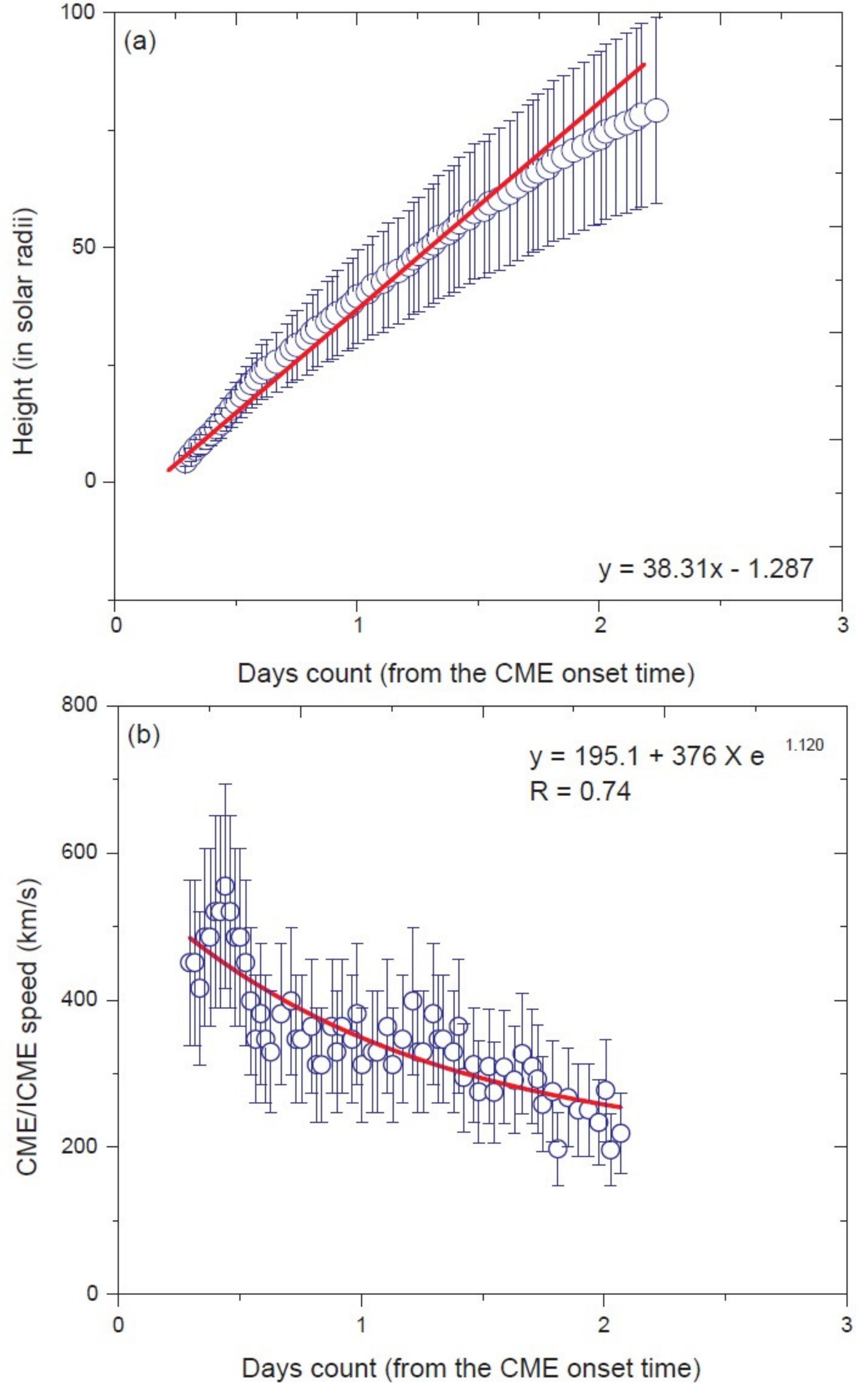}
      \caption{(a) Obtained CME/ICME height for various time from the STEREO J-map constructions (see Figure \ref{stereo_CME_CRF2}g). (b) Calculated CME/ICME speed in the STEREO FOV. These figures are clearly showing that the CME/ICME attained the solar wind speed in the interplanetary medium.}
\label{two_plots_CRF2}
\end{figure*}

We constructed elongation degree map based on the method developed by \cite{Sheeley1999} and \cite{Davies2009}. The inclined bright features in the J-maps (Figure \ref{stereo_CME_CRF2}g) correspond to the enhanced density structure of the CMEs erupted during 2-8 May 2013. We tracked the CME/ICME (marked by cross symbol) up to $\sim$17.50 elongation degree, equivalent to $\sim$80\(R_\odot\). We analyzed CME/ICME height-time information obtained from multi-point and multi-instrument observations. The calculated CME height (in \(R_\odot\)) and the speed values are given in the Figure \ref{two_plots_CRF2}a and b. From these figures, we understood that the main CME took approximately 5 days to reach the Earth environment. The average speed of the CME in the interplanetary medium is varying between $\sim$200 to $\sim$500 km $s^{-1}$.

\begin{figure*}
\centering
\includegraphics[width=0.75\textwidth]{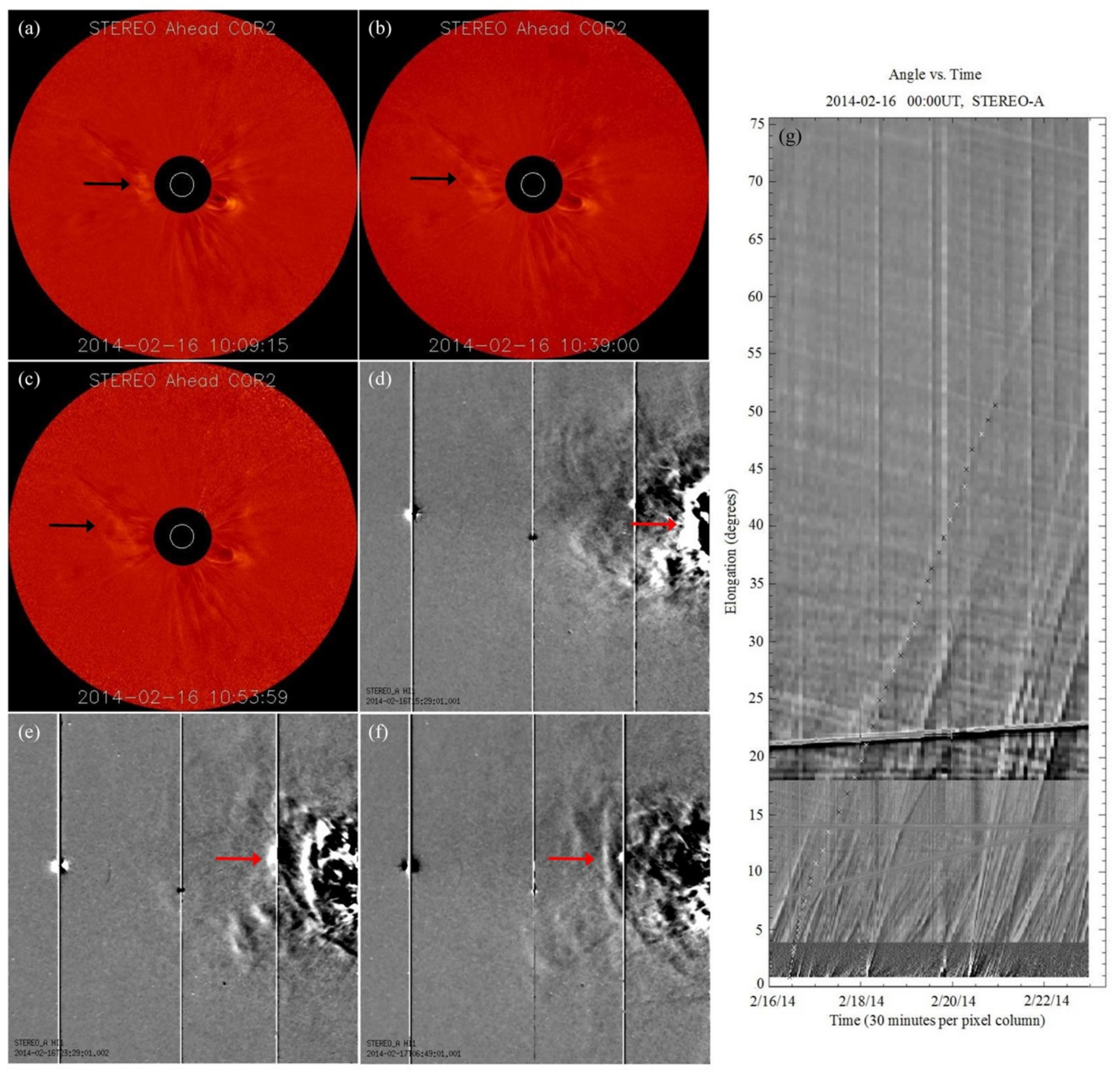}
\caption{CME/ICME observation in the STEREO FOV (for the corresponding CME see Figure \ref{LASCO_CME_CRF2}d and e). The evolution of the CME in (panels (a), (b) and (c)) COR2 and the same CME in (panels (d), (e) and (f)) HI1  FOV. The propagation of the CME/ICME is indicated by arrows. Time-elongation map (J-map) construction using the STEREO/SECCHI spacecraft observations during the interval of 16 to 22 February 2014.} 
\label{stereo_CME}
\end{figure*}

\begin{figure*}
  \centering
  \includegraphics[width=0.7\textwidth]{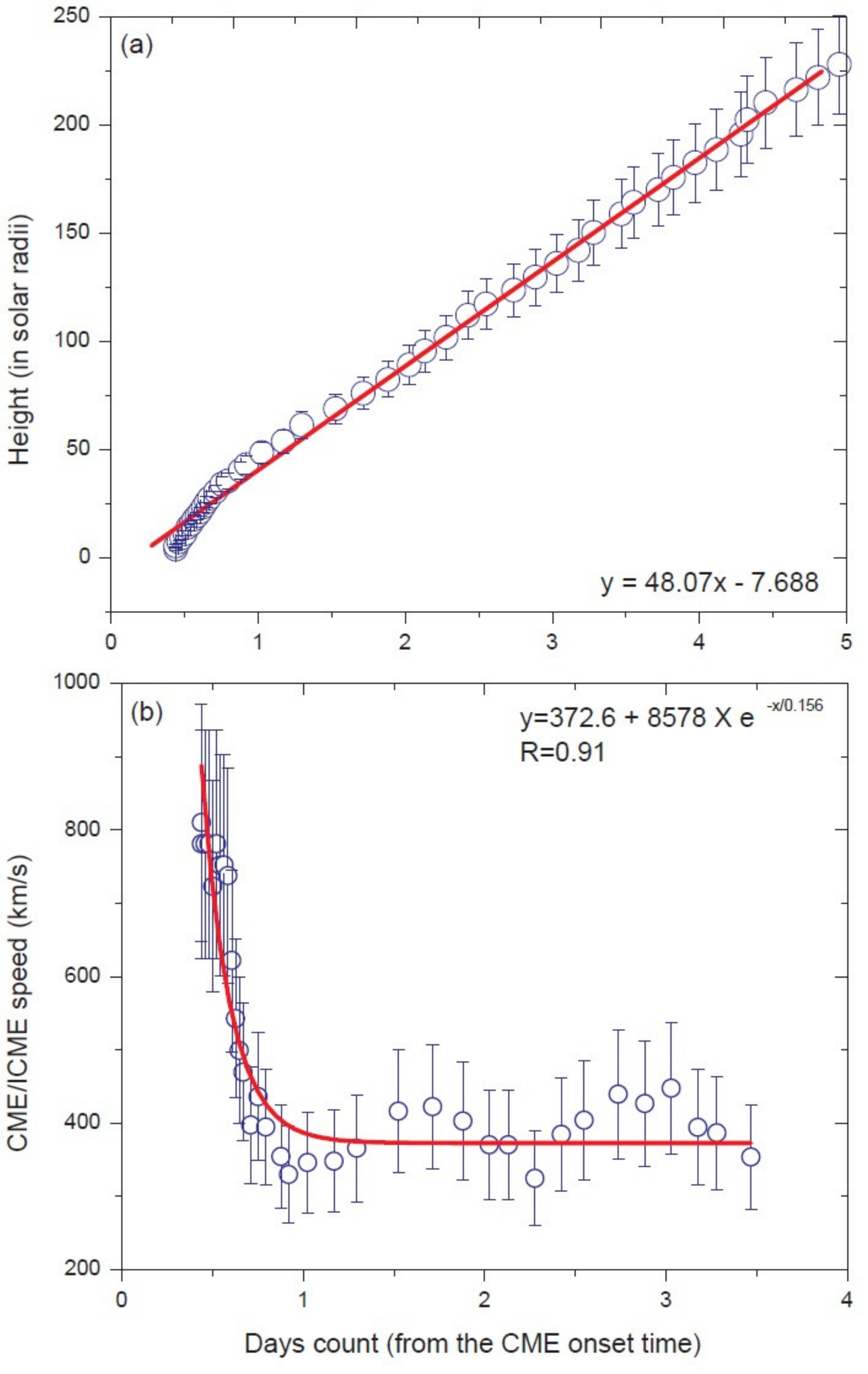}
       \caption{(a) Obtained CME/ICME height for various time from the STEREO J-map constructions (see Figure \ref{stereo_CME}g). (b) Calculated CME/ICME speed in the STEREO FOV. These figures are showing that the major geo-effective CME/ICME was propagating with the constant speed in the interplanetary medium.}
\label{two_plots}
\end{figure*}

Initial positions of the second event are indicated by the black arrow mark in STA COR1 FOV (see Figure \ref{stereo_CME}a and b). This particular CME observed in the STA around 10:09 UT. CME/ICME propagation is shown in Figure \ref{stereo_CME}c, d, e and f by white arrow. Corresponding J-map is shown in Figure \ref{stereo_CME}g. We tracked the geo-effective CME (marked by cross mark) up to $\sim$50.63 elongation degree, equivalent to $\sim$228\(R_\odot\). The calculated CME height (in \(R_\odot\)) and the speed are given in the Figure \ref{two_plots}a and b. The second geo-effective CME took approximately 4 days to reach the Earth environment. As noted earlier, this CME produced a major goe-effective event with the Dst value of -119 nT. Finally, this CME attains the solar wind speed in the interplanetary medium and its speed ranges between $\sim$350 to $\sim$420 km $s^{-1}$. 

\subsection{Signature of near Earth ICMEs}
\begin{figure*}
\centering
\includegraphics[width=0.75\textwidth]{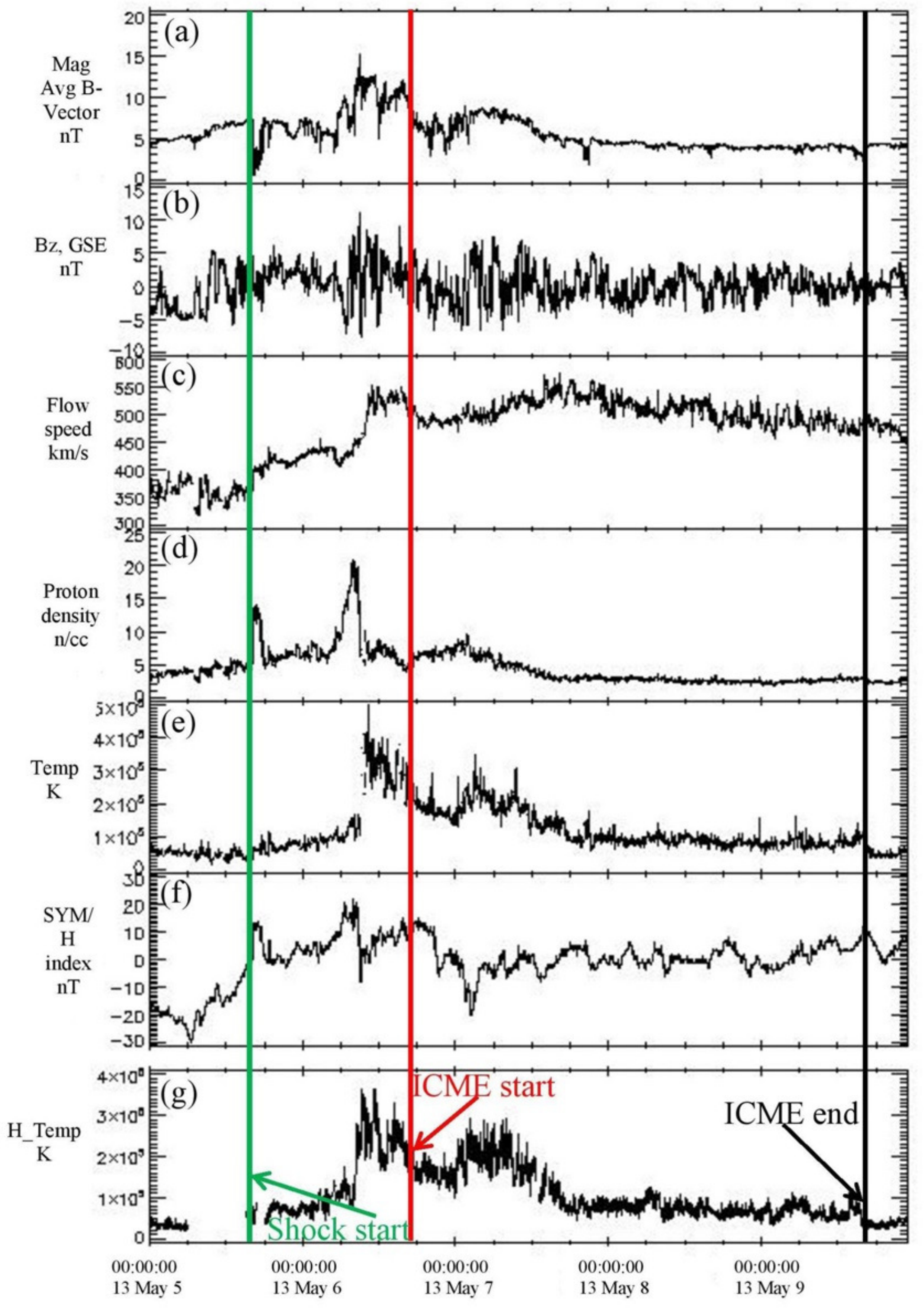}
\caption{Near Earth ICME signatures observed at 1AU by ACE spacecraft. These in-situ observations of the interplanetary CME corresponds to the CME shown in Figure \ref{LASCO_CME_CRF2}a and b. The vertical green, red and black lines denote the arrival of the IP shock, ICME start and end respectively.} 
\label{OMNI_CRF2}
\end{figure*}

We analyzed the in-situ observations taken from the Advanced Composition Explorer (ACE, located at L1 point) spacecraft to identify the ICME structure. Figure \ref{OMNI_CRF2} shows the variation of magnetic field and solar wind plasma parameters during 00:00 UT on 2013 May 05 to 23:59 UT on 2013 May 09. The arrival time of a interplanetary (IP) shock is indicated by a sudden enhancement in average magnetic filed, Earthward magnetic filed components, flow speed, proton density, temperature, and radial component of proton temperature and it was observed at 15:10 UT on 2013 May 05 (marked by the vertical green line). The region between the first and second vertical lines represents the turbulent sheath region. The arrival time of the ICME is indicated by low proton temperature and it is marked by the vertical red line around 16:45 UT on 2013 May 06. The observed arrival times of the IP shock and ICME are 81.77 hours and 107.35 hours, respectively. Also, the ICME ending time is marked by the vertical black line. We note a jump in the parameters which are associated with weak IP shock arrival. This weak shock may be related to the flank region of the CME, resulting in an insignificant impact in the near Earth region. Just before the shock arrival, we can see the minor fluctuation in the solar wind parameters which are probably caused due to a CIR originating from the coronal hole in the solar source regions.

\begin{figure*}
\centering
\includegraphics[width=0.75\textwidth]{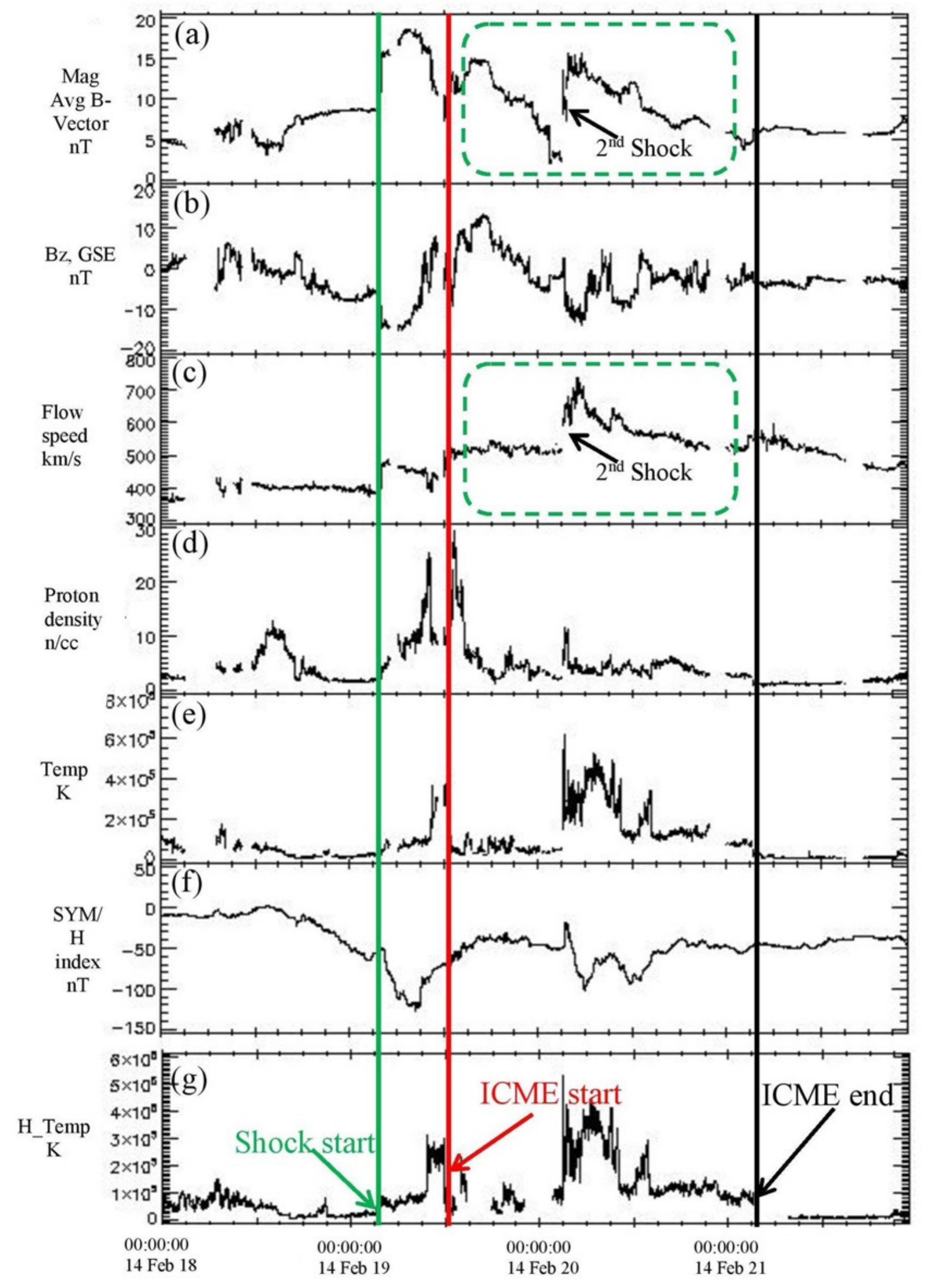}
\caption{Near Earth ICME signatures observed at 1AU by ACE space craft. These in-situ observations of the interplanetary CME corresponds to the CME shown in Figure \ref{LASCO_CME_CRF2}d and e. The vertical green, red and black lines denote the arrival of the IP shock, ICME start and end respectively. Interaction region of the ICMEs are noted in the green boxes. Another unknown IP shock clearly visible in this interaction region.} 
\label{OMNI}
\end{figure*}

Similarly, for second event variations of plasma and magnetic field parameters are given in Figure \ref{OMNI}. These parameters observed during 00:00 UT on 18 February 2014 to 23:59 UT on 21 February 2014. This event is a major geo-effective events and the beginning of storm disturbance is observed at 03:57 UT on 19 February 2014 (marked by the vertical green line). Vertical red line around 12:00 UT on 19 February 2014 indicates the ICME arrival time and the vertical solid black line is indicating the ICME ending time. Notably, the SYM/H Dst index is more negative during the interval of passage of the shock sheath region. The Dst index falls to a minimum value of -119~nT at $\sim$09:00~UT on 2014 February 19. Afterwards, Dst undergoes the recovery phase, but it recovered fully after six days. Before that Dst maintain the average value of -50~nT during 19-24 February 2014. The observed arrival times of the IP shock and ICME are 65.95 hours and 74.00 hours, respectively.

\subsection{Estimated Arrival Time using DBM and ESA models}
We estimated the arrival time of the ICME/IP shocks at 1 AU using the following CME/shock prediction model (a) Drag Based Model (DBM) \citep{Vrsnak2013}, and (b) Empirical shock Arrival model (ESA) \citep{Gopal2001}. These models are based on the near-Sun observations of the CME within the SOHO/LASCO FOV. The input parameters of the DBM models are: (1) first observational distance ($R_{o}$) of CME, (2) linear speed of the CME, (3) drag parameter ($\gamma$), and (4) asymptotic solar wind speed ($w$). The first observation point of the CME by C2 coronagraph is at a distance of 2.66~$R_{\odot}$ at 05:24~UT. Also, the linear speed of the CME in the LASCO FOV is around 671 km $s^{-1}$ and the solar wind speed (obtained from the ACE at 1AU) is around 400 km $s^{-1}$. Using these initial input parameters and DBM model, we calculated the CME transit time. The time taken by the CME from its first observation on the Sun to near-Earth is nearly 72.05 hours with impact speed (at 1 AU) of~513 km~s$^{-1}$. Further we calculated the arrival time of the CME using Empirical Shock Arrival (ESA) model which is briefly explained in the previous paper \citep{Syed2015}. The estimated arrival time for the IP shock is 84.04 hours. Here, we used the following acceleration equation a=2.19-(0.0054*CME linear speed) for arrival time which is based on the work of \cite{Michalek2004}. Summary of the observations and model predictions are given in the Table \ref{Tab:cme_summary}. From this comparison, we also confirm the CME-ICME connection.

\begin{table}
\caption{Summary of the first CME initial observations, along with the actual and estimated transit times.}
\begin{tabular}{ll}
\hline
First observation in LASCO FOV & 05:24 UT on 2013 May 02 \\
Height of first observation in LASCO FOV & 2.66 R$_\odot$\\
Linear speed (from LASCO FOV observations)& 671~km~s$^{-1}$ \\
IP Shock arrival at Near-Earth & $\sim$15:10 UT on 05 May 2013\\
ICME arrival at Near-Earth & $\sim$16:45 UT on 06 May 2013\\
Actual transit time for IP shock & \\
~~~~~from the first point of the CME observation & 81.77 hours\\
Actual Transit time for ICME  &\\
~~~~~from the first point of the CME observation & 107.35 hours\\
Estimated transit time &\\
~~~~~from DBM & 72.05 hours\\
Estimated transit time &\\ 
~~~~~from ESA model & 84.04 hours\\
\hline
\end{tabular}
\label{Tab:cme_summary}
\end{table}

\begin{table}
\caption{Summary of the CME initial observations, along with the actual and estimated transit times.}
\begin{tabular}{ll}
\hline
First observation in LASCO FOV & 10:00 UT on 2014 February 16 \\
Height of first observation in LASCO FOV & 2.55 R$_\odot$\\
Linear speed (from LASCO FOV observations)& 634~km~s$^{-1}$ \\
IP Shock arrival at Near-Earth & $\sim$03:57 UT on 19 February 2014\\
ICME arrival at Near-Earth & $\sim$12:00 UT on 19 February 2014\\
Actual transit time for IP shock & \\
~~~~~from the first point of the CME observation & 65.95 hours\\
Actual Transit time for ICME  &\\
~~~~~from the first point of the CME observation & 74.00 hours\\
Estimated transit time &\\
~~~~~from DBM & 74.38 hours\\
Estimated transit time &\\ 
~~~~~from ESA model & 64.47 hours\\
\hline
\end{tabular}
\label{Tab:cme_summary_2}
\end{table}

The first detection point of the second CME by C2 coronagraph is at a distance of 2.55~$R_{\odot}$ at 10:00~UT. Also, the linear speed of the CME is around 634 km $s^{-1}$ and the solar wind speed is around 400 km $s^{-1}$. Using the initial parameters and DBM model, we calculated the CME transit time. The time taken by the CME from the Sun to near-Earth is nearly 74.38 hours with near Earth speed (at 1 AU) of~504 km~s$^{-1}$. Then using the Empirical Shock Arrival (ESA) model, we estimated arrival time for the IP shock to be about 64.47 hours. Comparison between the observations and model predictions are given in the Table \ref{Tab:cme_summary_2}. From the comparison, we confirm the CME-ICME connection. For both the cases, IP shock arrival showing good match with the actual observations. In the first case, the prediction error is
little high between the estimated and actual ICME arrival time which could be attributed to the deflection of the erupting structure and/or interaction between the two propagating structures of CME/ICME in the near-Sun region or IP medium.

\section{Summary and Conclusions}
\label{sec:conclusions}
In this article, we have carried out a study of two Earth directed CMEs. We connect and correlate the flare, CME and ICME information from the near-Sun region to the near-Earth environment. In the following, we summarize the important results obtained from the present study.  

Both the events clearly show the pre-existence of a small twisted filament associated with the eruptive circular ribbon flares. The characteristics of CRF are briefly discussed from multi-channel AIA observations and available H$\alpha$ images. The formation of circular ribbons at the source regions of both the CMEs indicate that the coronal magnetic field configuration during both the events are similar. It is well known that CRFs are associated with typical fan-spine configuration (see e.g., \citep{Pooja2020}. Both the CRFs initiated from complex magnetic field configuration of the active region that exhibit $\beta\gamma$ and $\beta\gamma\delta$ sunspot distribution, respectively, for the first and second event. At the time of flares, e-CALLISTO radio spectrometer observed strong type III radio burst. Despite the similarities in the source regions configuration and some of the flare characteristics, both CME events produce quite different geo-effective behavior. First event is also associated with type II radio burst \citep{Gopalswamy2001,Lug2017}. Notably, during the first event, we observed CME-CME interactions which was absent during the second event. CME-CME interaction and associated density perturbation increase the probability of type II formation which is likely the case for the first event reported here. From metric type II radio observation and considering Newkirk density model \citep{Newkirt1961}, we estimate the shock formation height range of $\approx$1.18 - 1.91R$_{\odot}$ with the corresponding shock speed (782 km $s^{-1}$). 

After the eruption of the flares, CMEs were observed in the LASCO C2 FOV with moderate linear speeds of 671 km $s^{-1}$ and 634 km $s^{-1}$ respectively. Further, the CMEs/ICMEs were observed by the STEREO instrument from Sun to 1AU distance. The in-situ measurements confirm CMEs associated IP shocks as well as the ICMEs at 1 AU. 

These two CMEs present some similar characteristics, eg. (i) Similar intensity class of associated flare (M1.1), (ii) circular ribbon flares, and (iii) moderate speed CMEs. However, these events differ in the way of (i) early evolution and propagation of CME during its activation phase in the corona, and (ii) near-Sun pre-post CME interaction. At near-Earth region, the main difference is that the second event is associated with the major geomagnetic storm with DST $\approx$-119 nT in comparison to the first event which is associated with the Dst of $\approx$-20 nT. This is the first detailed study about the geo-effectiveness of circular ribbon flares and flare-CME-ICME connections.

The following points are found to be the reasons for the major 
geo-effectiveness of the second CME:

\begin{itemize}

\item {While the second CME erupted from the exact disk center of the Sun, the first CME erupted 25$^{\circ}$ away from the solar disk center. The location of CME in the solar disk is an important factor that decides its trajectory in the corona and subsequent evolution in the interplanetary medium \citep{Zhang2003, Manoharan2004, Dasso2007, Zhang2011}. Also, when compared to the first event, the second CME was associated with a larger filament eruption (for example see the study by \cite{Chandra2017}.}

\item {From the AIA difference images (see Figure \ref{Deflection}), we found that the first eruption is highly deflected from the source active region. Also the eruption direction is not along the radial direction. After the main eruption, a huge part of the CME moved towards the North-East direction of the Sun. The non-radial evolution of CME as it moves outward from the source active has been reported in earlier studies \citep[see e.g.,][]{Zukka2017, Prabir2020}. Finally, only a small part of the CME propagates along the Sun-Earth line. Notably, there was no deflection in the case of the second CME, so that most of the CME portion reached the Earth.}

\item {First CME seems to be interacted with two pre-CMEs in the LASCO FOV. These two pre-CMEs erupted consecutively  around 01:25 UT and 04:24 UT. Very likely, because of the interaction, the first CME might have lost its energy \citep{Gopalswamy2001, Yash2014, Shan2014, Joshi2018, Moro2020, Rod2020, Scol2020}.}

\item {Our analysis also suggests that the first event got dissipated in the interplanetary space. After a certain distance, the first ICME mingled up with the solar wind.}

In summary, the present papers provides a detailed multi-wavelength, multi-instrument, and multi-point observations of two moderate speed CMEs, both of which were capable to reach the near-Earth region. The two events seems to be identical in terms of morphology of associated flare and origin of CME at the source region by filament eruption. The source region of both the events are also in the central part of the solar disk but within a separation of 25$^{\circ}$. However, both CMEs evolve differently beyond the lower coronal region. Our study shows that the CME deflection by the large-scale coronal structures and CME-CME interactions were the major reasons that largely shaped the prorogation characteristics of the two events in the upper coronal and interplanetary medium. The present investigation points toward the importance in detecting the changes in the prorogation characteristics of the CMEs through a combination of multi-channel and multi-point measurements, such as, source region imaging, radio spectral diagnostic, heliospheric imaging, and in-situ observations. The understanding of the Sun-Earth propagation characteristics of CMEs is a key toward assessing their geo-effective behavior. We plan to analyze the CME observations of a series of circular ribbon eruptive flares in future.

\end{itemize}

{\bf{Acknowledgment}}
We thank SDO, CALLISTO, and GONG teams for their open data
policy. We sincerely thanking Aryabhatta Research Institute of Observational Sciences (ARIES) solar observation staffs. We are grateful to the Solar Geophysical Data team, the World Data Center for Geomagnetism (Kyoto University), and the OMNIWeb Plus data and service for their open data policy. The CME catalog used in this study is generated and maintained by the Center for Solar Physics and Space Weather, The Catholic University of America, in cooperation with the Naval Research Laboratory and NASA. The data services from CDAWeb are also thankfully acknowledged.


\end{document}